%% file: main.tex
\documentclass[conference]{IEEEtran}
\IEEEoverridecommandlockouts

\usepackage{array}
\usepackage{multirow}
\usepackage{threeparttable}

\usepackage{cite}
\usepackage{amsmath,amssymb,amsfonts}
\usepackage{algorithmic}
\usepackage{graphicx}
\usepackage{bmpsize}
\usepackage{xcolor}
\usepackage{lipsum}
\usepackage[colorlinks=true,urlcolor=black]{hyperref}
\def\BibTeX{{\rm B\kern-.05em{\sc i\kern-.025em b}\kern-.08em
    T\kern-.1667em\lower.7ex\hbox{E}\kern-.125emX}}
\usepackage{float}

\newcommand{\prune}{\textsc{PRUNE}}

\newcommand{\minimize}{\textsc{MINIMIZE}}

\newcommand{\slack}{\textsc{SLACK}}

\begin{document}
\hyphenation{Block-chain Trans-ac-tions}

\title{On the Storage Overhead of Proof-of-Work Blockchains
\thanks{This work was supported in part by the by the European Commission H2020 TeraFlow Project under Grant Agreement No 101015857.}
}

\author{\IEEEauthorblockN{Alessandro Sforzin}
\IEEEauthorblockA{NEC Laboratories Europe}
\and
\IEEEauthorblockN{Matteo Maso}
\IEEEauthorblockA{AiSight}
\and
\IEEEauthorblockN{Claudio Soriente}
\IEEEauthorblockA{NEC Laboratories Europe}
\and
\IEEEauthorblockN{Ghassan Karame}
\IEEEauthorblockA{Ruhr-University Bochum}
}

\maketitle
\begin{abstract}
Permissionless blockchains such as Bitcoin have long been criticized for their high computational and storage overhead. Unfortunately, while a number of proposals address the energy consumption of existing Proof-of-Work deployments, little attention has been given so far to remedy the storage overhead incurred by those
blockchains. In fact, it seems widely acceptable that full nodes supporting the blockchains have to volunteer hundreds of GBs of their storage, to store and verify all transactions exchanged in the system.

In this paper, we explore the solution space to effectively reduce the storage footprint of Proof-of-Work based blockchains. To do so, we analyze, by means of thorough empirical measurements, how existing full blockchain nodes utilize data from the shared ledger to validate incoming transactions/blocks. Based on this analysis, we show that it is possible for full nodes to locally reduce their storage footprint to approximately 15 GB, without any modification to the underlying protocol. We also discuss other client-side strategies to further reduce the storage footprint while incurring negligible computational overhead on the nodes.
\end{abstract}

\input{introduction.tex}

\input{background.tex}

\input{analysis.tex}

\input{methodology.tex}

\input{evaluation.tex}

\input{conclusions.tex}

\bibliographystyle{abbrv}
\bibliography{bibliography}

\end{document}

%% file: introduction.tex
\section{Introduction}
\label{sec:intro}

Blockchains are receiving increasing attention among researchers and practitioners, owing to their promise to efficiently manage business processes in a decentralized manner. Although the literature features a large number of blockchains that notably differ in their leader election and consensus protocols, almost all existing blockchains rely on the presence of a shared ledger that enables blockchain nodes to agree on the order and correctness of information (i.e., transactions and blocks).

Permissionless blockchains (such as Bitcoin and Ethereum) have been heavily criticized due to their high computational and storage overhead. At the time of writing, Bitcoin---arguably the most popular instantiation of a permissionless blockchain---incurs an estimated annual energy consumption of 105.70TWh (see ~\cite{cbeci}), and needs more than 370~GB of space to store the ledger.

While the computational overhead in existing blockchains can be remedied by replacing Proof-of-Work with newer, more energy-friendly protocols such as Proof-of-Stake or Byzantine-Fault-Tolerant protocols, little work exists to remedy the storage overhead incurred by today's blockchains.

A high storage overhead is mostly evident in those blockchains that have witnessed the largest adoption. That is, the larger is the adoption of a given blockchain platform, the more transactions that are exchanged, and in turn the bigger is the storage required to maintain the shared ledger.
On the one side, keeping all transaction data is essential to ensure the security of the system and to make sure that no transaction is spent more than once. On the other side, storing hundreds of GBs, is one of the main reasons why many users shy away from running full nodes (i.e., nodes that store the full ledger).

Previous work has proposed a number of fixes and extensions to mitigate the storage overhead of existing blockchain platforms~\cite{palm2018selective,chepurnoy2016rollerchain,marsalek19,matzuttKPDHW20}. Most proposals introduce protocols to create and maintain periodic checkpoints (or snapshot) of the ledger, so that data belonging to previous snapshots can be safely deleted. However, implementing a checkpointing strategy require either to modify the way the blockchain works or a fork.

In terms of deployed solutions, Bitcoin allows users to prune
the blockchain storage by defining a threshold (in GB or in block height) below which content in their local copy is trimmed~\cite{BIP_prune}. Such pruning techniques are available in Bitcoin clients and can be locally used as a stand-alone, independent solution to reduce the storage overhead of the blockchain. However, the choice of the threshold value (either in terms of disk space or in terms of block height) is left to the user without any guidelines. On the one hand, storing too many blocks may not be feasible for all clients and  would cause unnecessary data to be stored on disk. On the other hand, storing too few blocks would result in the deletion of data that may still be necessary to verify unspent transactions. Note that verification of transactions for which data has been deleted, incurs in additional communication overhead to fetch the required data from the network.

In this paper, we address the problem of efficiently managing the storage overhead incurred by existing blockchains and present the first study on how blockchain nodes use data from the shared ledger to validate transactions and blocks. We focus on Bitcoin since (i) it is arguably the most popular blockchain where the storage problem is most relevant, and (ii) publicly available data allows us to study the behavior of the network over a fairly long time period and obtain meaningful insights towards reducing the storage footprint. To do so, we start by analyzing, by means of thorough empirical measurements, how existing Bitcoin nodes manage data from the shared ledger to validate incoming transactions/blocks. Based on these findings, we explore the solution space to effectively manage the storage of existing PoW blockchains. To this end, we adapted a blockchain parser based on~\cite{py_parser} to compute the storage savings of the various strategies we devise.

Unlike common beliefs, our results show that it is possible for full Bitcoin nodes to locally reduce their storage footprint by approximately 95.9\% without any modification to the underlying protocol and with no appreciable overhead to validate transactions. Moreover, we show that an archival node---wishing to store all information in the blockchain without any loss---could save up to 29\% of storage space without losing any information from the ledger. This results in 5-10\% more storage savings when compared to existing compression algorithms that can achieve a maximum compression rate of up to 24\% on the Bitcoin ledger---and without requiring the heavy computational load associated with (de-)compression. Our parser will be released as open-source to better aid the community in estimating the actual storage needs of Bitcoin nodes as the ledger grows in size. We stress at this point that our observations are not restricted to  Bitcoin and equally apply to the myriad of altcoins (or forks of the Bitcoin blockchain) that are currently deployed (e.g., Dogecoin, Bitcoin Cash, Litecoin, Monacoin).

The remainder of the paper is organized as follows.
In Section~\ref{sec:back}, we overview the storage requirements in existing blockchains and discuss related work in the area. 
In Section~\ref{sec:analysis}, we empirically measure the transaction age and the storage overhead incurred in current Bitcoin transactions. 
In Section~\ref{sec:meth}, we explore the
space of possible solutions that may allow a Bitcoin node to reduce
the ledger’s footprint on its local storage and we evaluate the effectivness of those strategies in Section~\ref{sec:eval}.
Finally, we conclude the paper in Section \ref{sec:concl}. 

%% file: background.tex
\section{Background \& Related Work}
\label{sec:back}
In this section, we introduce relevant background on Proof-of-Work (PoW) blockchains, with a focus on transactions validation and storage.

\subsection{The Need for Storage in Existing Blockchains}
PoW-based blockchains leverage Proofs of Work (PoW) as a public timestamping mechanism in order to prevent double-spending attacks. In practice, transactions are broadcasted and special nodes called \emph{miners} add those transactions that they consider valid in a so-called \emph{block}. A block is valid only if it contains valid transactions and the solution to a cryptographic puzzle. New blocks are cryptographically tied to previous ones via hash chains and, even if different chains of blocks can co-exist, only the longest chain is considered valid. In a nutshell, only transactions included in the blocks of the longest chain are considered valid.
A miner that outputs a new block, broadcasts it so that other nodes can check its validity by checking the validity of its transactions and the correctness of the solution to the cryptographic puzzle. Nodes that are not mining (i.e., that do not contribute to block creation) are called full nodes. These nodes verify all exchanged information (blocks and transactions) in the blockchain and therefore have to store the full blockchain ledger\footnote{Without this information, an adversary can perform history corruption attacks---effectively presenting another chain of blocks (and the transactions therein) as the ``main'' chain.}---albeit without any explicit incentives.

For example, at the time of writing, a miner that succeeds in mining a block receives a fixed revenue of 6.25 BTCs and a variable profit comprising of all the fees that are included in the confirmed transactions.
Full nodes are not rewarded by any means in spite of their critical role to preserve the security of the whole system.

\begin{figure}[t]
	\centering
	\includegraphics[width=0.7\columnwidth]{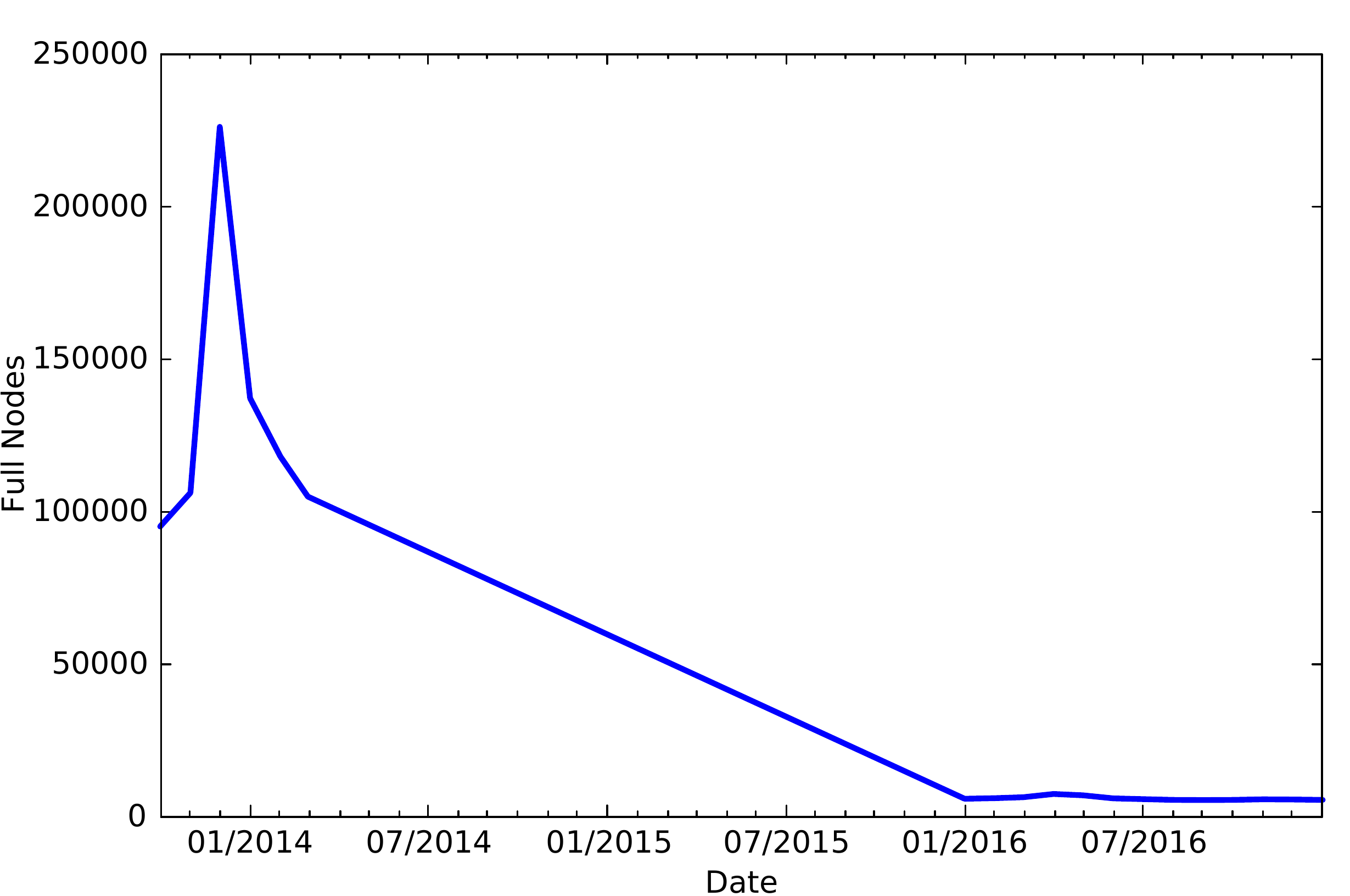}
	\caption{Evolution of the number of Bitcoin full nodes over time. Estimates are adapted from~\cite{full2}.}
\vspace{-1.5 em}
	\label{fig:2}
\end{figure}

Given the huge adoption of PoW-based blockchains~\footnote{For instance, Bitcoin processes around 2000 transactions every 10 minutes at the time of writing.}, the storage requirements on full nodes has considerably increased.
For instance, in 2014, the Bitcoin ledger was approximately 15 GBs. Throughout 2021, Bitcoin's ledger grew to approximately 371 GBs. In turn, as shown in Figure~\ref{fig:2}, the number of full Bitcoin nodes (that store the full ledger) dropped from 200,000 in 2014 to approximately 5,000 in 2016. Given lack of incentives, this number is only expected to decrease in the future.

We note that the majority of blockchain platforms also allow nodes that act as ``lightweight clients''. A lightweight client---usually a device with limited resources such as a smartphone---only downloads and verifies a small part of the chain. For example, the Bitcoin community provides the BitcoinJ\footnote{\url{https://bitcoinj.org/}}, PicoCoin\footnote{\url{https://github.com/jgarzik/picocoin}}   and Electrum\footnote{\url{https://electrum.org/}} clients implementing the Simple Payment Verification (SPV) mode~\cite{whitepaper}, where the clients connect to a full node that has access to the complete blockchain and can help the client to confirm transactions. While lightweight clients ease the adoption of blockchain technology, we argue that full nodes are essential for a blockchain platform to thrive and solutions to minimize the burden (including the storage burden) of full nodes are necessary.

\subsection{Transactions \& Scripts}
\label{sub:btc}
Many permissionless blockchains (such as Bitcoin) rely on the concept of Unspent Transaction Output (UTXO) to capture transactions in the platform.
Essentially, a transaction is defined by a set of inputs (TxIn) and outputs (TxOut) that dictate the transfer the ownership of a sets of coins, specified in the TxIns, to a set of Bitcoin addresses, each of which get the amount of coins specified in the TxOuts. Concretely, each TxIn refers to one or more TxOuts of older transactions.

A transaction is invalid if the value of its TxOuts exceeds the value of its TxIns. However, if the TxIns value exceeds the value of the TxOuts, the miner who outputs the block storing the transaction can claim the difference as a transaction fee. The exact conditions under which an output can be spent are encoded with a set of scripts, and only the participants that are able to provide the correct input to the script, such that it evaluates to true upon execution, are allowed to spend the coins output by a given Bitcoin transaction.

Scripts refer to a custom non-Turing complete scripting language that are designed with the aim to support different types of transactions and extend the applicability of transaction beyond the simple transfer of funds. Scripts are stack-based, support a number of functions (commonly referred to as opcodes), and either evaluate to true or false. The language supports dozens of different opcodes ranging from simple comparison opcodes to cryptographic hash functions and signature verification. Since scripts are supposed to be executed by all blockchain node, they could be abused to conduct denial-of-service attacks; therefore, a considerable number of opcodes have been temporarily disabled. This was one of the main reasons why scripts do not provide
rich support when compared to standard programming languages. {The most common type of scripts found in Bitcoin's historical data are \emph{pay-to-pubkey-hash} (P2PKH), \emph{pay-to-pubkey} (P2PK), and \emph{pay-to-script-hash} (P2SH).

Figure~\ref{fig:3} depicts a simplified transaction with one input and two outputs. In this example, the transaction spends $w$ BTCs to address $X$ and $x$ BTCs to address $Y$. The outputs that have not yet been spent (i.e., the two outputs of the transaction), are commonly referred to as unspent transaction outputs (UTXO).

Bitcoin keeps an up-to-date database of UTXOs, which it updates by adding or removing TxOuts created, or spent, by new transactions. 

\begin{figure}[t]
	\centering
	\includegraphics[width=0.7\columnwidth]{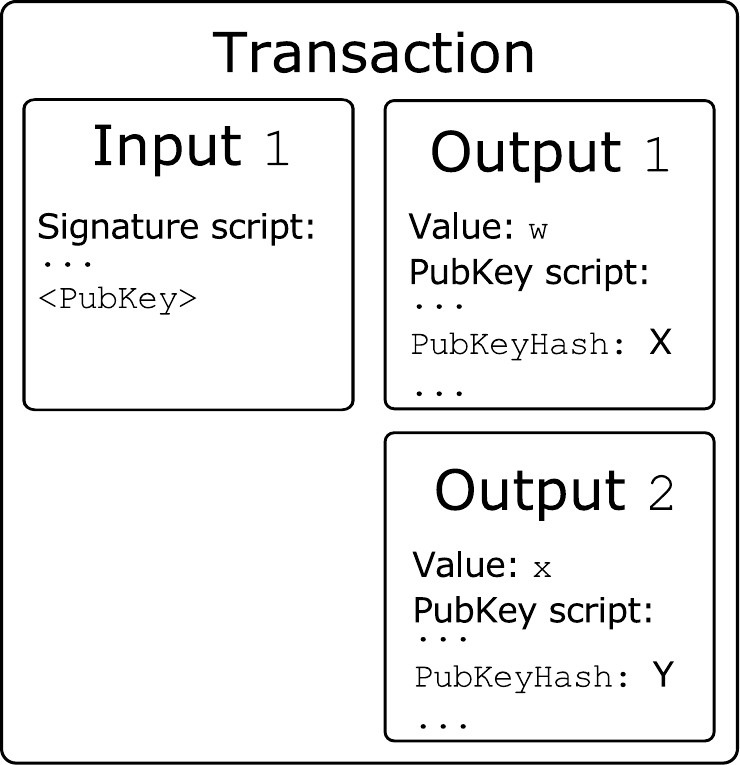}
	\caption{Example of a Bitcoin transaction with 1 input and 2 outputs.}
\vspace{-1.5 em}
	\label{fig:3}
\end{figure}

\subsection{Existing methods to shrink storage}

As mentioned earlier, little work has addressed the problem of reducing the storage footprint of blockchain nodes..

Florian et al.,~\cite{florian2019erasing} suggest that UTXOs can be deleted and space can be saved if one is willing to trust other nodes to verify the validity of transactions including UTXOs that have been locally erased---essentially, the node acts as a lightweight client if the transaction to be validated includes at least one UTXO that has been deleted.

A number of proposals suggest to limit the size of the data to be stored by using an account-based model where the system keeps track only of those accounts that have a positive balance~\cite{palm2018selective,bruce2014mini}.
Other proposals~\cite{chepurnoy2016rollerchain,marsalek19,matzuttKPDHW20,bruce2014mini} introduce extensions to existing systems to create and maintain periodic snapshots (or checkpoints) of the ledger, so that data belonging to previous snapshots can be safely deleted.

Real-world solutions focus on ``pruning''. For instance, pruning in Bitcoin (and other altcoins) was introduced with Bitcoin Core v0.11~\cite{BIP_prune} in 2015. Concretely, nodes can set a flag to specify the amount of disk space that Bitcoin can use for blocks and data, starting from a minimum of 550 MB (288 blocks, about two days worth of blocks). However, the choice of the threshold value
(either in terms of disk space or in terms of block height) is left
to the user without any guidelines. On the one hand, storing
too many blocks may not be feasible for all clients and would
cause unnecessary data to be stored on disk. On the other
hand, storing too few blocks would result in the deletion of
data that may still be necessary to verify unspent transactions. Such nodes can neither relay missing blocks, nor maintain transaction indexes anymore.
Note that verification of transactions for which data has been
deleted, incurs in additional communication overhead to fetch
the required data from the network.

More recently, blockchains such as Bitcoin introduced the segregated witness (segwit) structure, which stores data required to validate transactions, that is, scripts and signatures, outside of the associated blocks~\cite{BIP_seg_wit}. While this solution reduces the communication overhead, it does not necessarily lead to a reduction in the storage overhead, since nodes must still store all the scripts/signatures that are relevant for the verification of unspent transactions.

%% file: analysis.tex
\section{Dynamics of Bitcoin Storage}
\label{sec:analysis}
In this section, we analyze the Bitcoin ledger in order to understand the underlying dynamics of blocks, UTXOs and, more in general, data utilization in the blockchain. Our goal is to extract insights that may aid the design of a storage-saving strategy.

We conducted these experiments by leveraging two existing open source tools, namely \texttt{bitcoin-blockchain-parser} \cite{py_parser} and \texttt{bitcoin-tools} \cite{btctools}, both written in Python.
The \texttt{bitcoin-blockchain-parser} parses Bitcoin's raw data stored on disk by Bitcoin's software (\texttt{bitcoind}).
We use it to scan Bitcoin's historical data (e.g., \texttt{blk*.dat} files) for the range of blocks to be analyzed, and for the entire blockchain to apply a storage optimization method.
The library \texttt{bitcoin-tools} parses Bitcoin's block index and chainstate.
The block index stores information for every block (e.g., block header and number of transactions in that block), and where each block is stored on disk.
The chainstate stores Bitcoin's current UTXO set.
Our tool leverages \texttt{bitcoin-tools} to fetch and decode the UTXO set. We used these parser to analyze the Bitcoin blockchain from the genesis block until block 684,816 (minted on May-25 2021).

\subsection{UTXO Distribution and Lifespan}

We start by looking at the ``current'' UTXO set (i.e., UTXO set at block 684,816) made of 123,394,434 UTXO in total. As shown in~\cite{dormant}, a very large number of UTXOs are ``dormant'', i.e., they have been confirmed early in the blockchain's history and not spent thus far. Namely, our measurements show that most of the blocks (84\%) confirm at least one of the current UTXOs. The first 105,000 blocks (i.e., between Jan-09 2009 and Jan-28 2011) account for 42,802 of the UTXOs in the current UTXO set. By examining the latest 25,000 blocks, we only find 642 UTXOs.

There may be multiple reasons why a UTXO has been created a long time ago and never been spent. One possible reason is that the owner is holding those UTXOs speculating that BTC value will increase. Other options are, e.g., dust UTXO or UTXO belonging to secret keys that have been lost. P{\'{e}}rez{-}Sol{\`{a}} et al.,~\cite{perez-sola} studied dust or unprofitable UTXOs in Bitcoin and found that, depending on the spending fee, up to 50\% UTXOs at block 507,964 (minted on Feb-6 2018) could be considered dust.

\begin{figure}[tbp]
    \centering
    \includegraphics[width=0.47\textwidth]{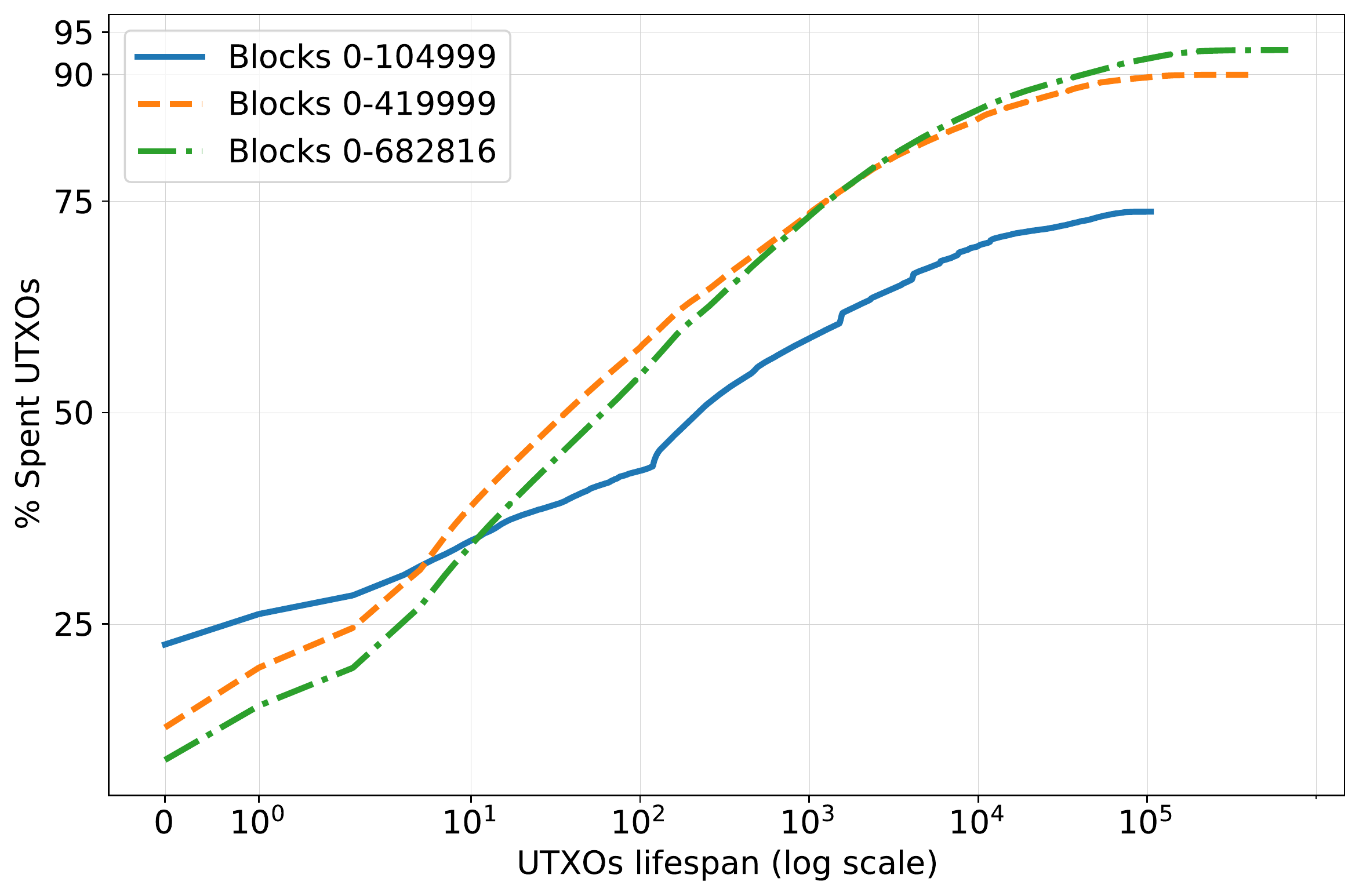}
     \vspace{-1 em}
    \caption{Lifespan (in blocks) of UTXOs created during the first 105,000, 420,000, and 682,816 blocks.}\label{fig:tx_life_cdf}
    \vspace{-1 em}
\end{figure}

In order to distinguish whether holding UTXOs for a very long time is common practice throughout the Bitcoin blockchain lifetime, we look at the lifespan of a UTXO, that is, the time---measured in blocks---between its addition to the blockchain (i.e., its parent block is added to the chain) and its expenditure (i.e., the block storing a transaction that spends the UTXO is added to the chain) at different time snapshots in the Bitcoin blockchain. Figure~\ref{fig:tx_life_cdf} shows the 50th, 90th, and 95th percentile of the UTXO lifespan across different time intervals. The first interval comprises the time from the genesis block until block 104,999, minted on Jan-28 2011: the blue curve of Figure~\ref{fig:tx_life_cdf} shows that less than 75\% of the UTXOs created within blocks 0-104,999 were actually spent before block 104,999. The second time interval goes from the genesis block until block 419,999, minted on Jul-09 2016: the corresponding dashed orange curve shows that UTXOs created within blocks 0-419,999 have a different behavior from the ones in the previous interval; roughly 90\% of those UTXOs have a lifespan smaller than 100,000 blocks. Finally, the third interval consider all blocks up to 684,816 (minted on May-25 2021): here we witness (green curve with dashes/dots) a trend similar to the one in the previous interval as 90\% of the UTXOs are spent within 42,000 blocks.

In a nutshell, Figure~\ref{fig:tx_life_cdf} shows that, albeit a large number of UTXOs are dormant---especially the ones created during the early years of Bitcoin---most of the UTXOs have a rather short lifetime. More precisely, we note that 90\% of all UTXOs are currently spent within 42,586 blocks (cf. Figure~\ref{fig:tx_life_cdf}). As Bitcoin becomes increasingly popular, we expect the UTXO lifespan to decrease over time.

\subsection{Data Distribution within Bitcoin Transactions}

Next, we focus on the various data types that are included in a Bitcoin transaction and compute the fraction of transaction data used to store the block header, transaction header, TxIns, TxOuts, and scripts. Here, we do not distinguish between the individual scripts---rather we measure deduplication level in scripts regardless of their type.
To take into account segregated witness data, we divide the analysis in two different time-periods: the period before the introduction of segwit, and the one after the introduction of segwit.

\begin{figure}[tbp]
    \centering
      \vspace{-1 em}
    \includegraphics[width=0.50\textwidth]{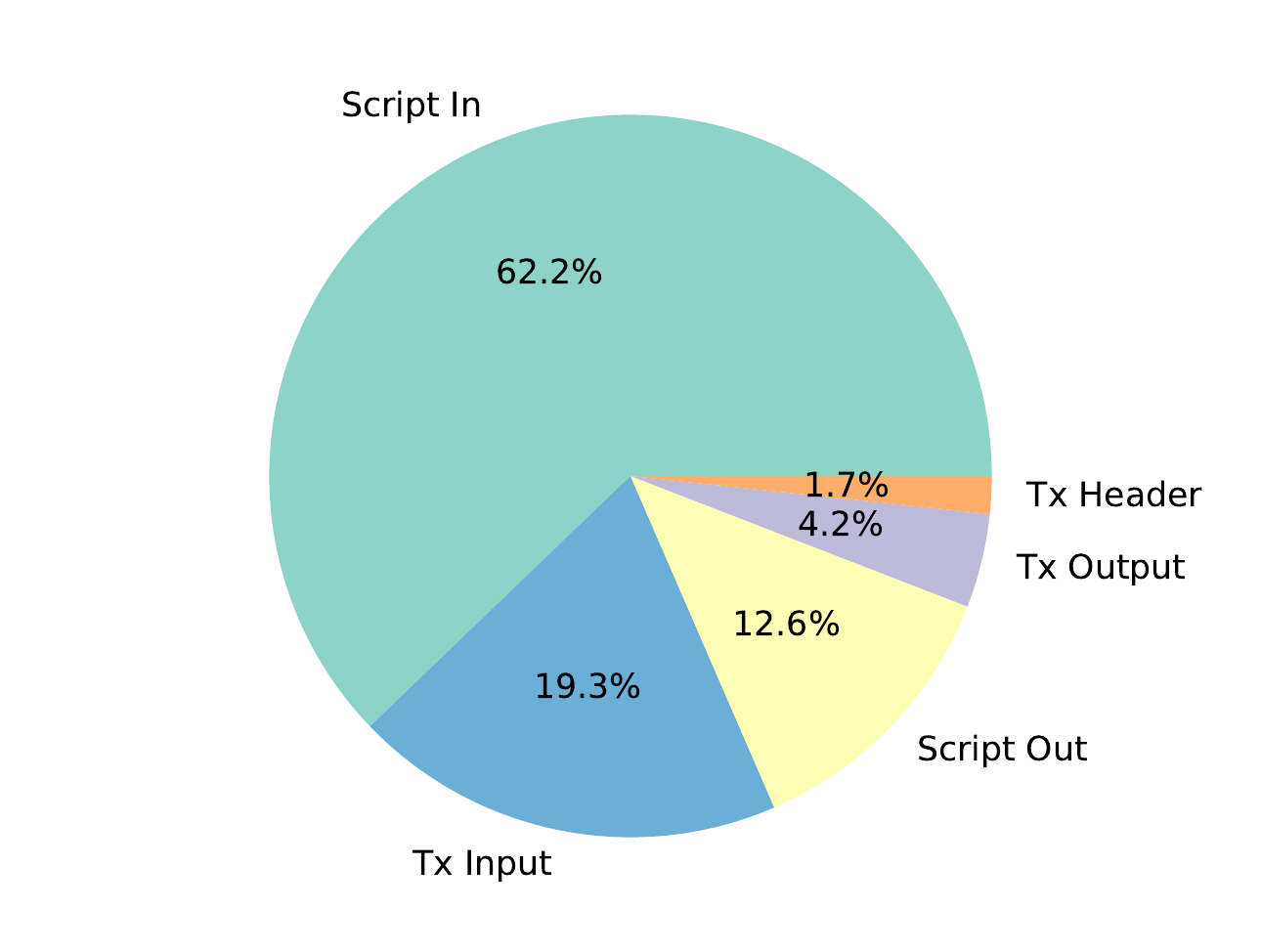}
      \vspace{-3 em}
    \caption{Distribution of data types within transactions from block 0 to block 481,823 (i.e., prior to the introduction of segwit).}\label{plot_no_segwit}
    \vspace{-1 em}
\end{figure}

\begin{figure}[t]
    \centering
      \vspace{-1 em}
    \includegraphics[width=0.50\textwidth]{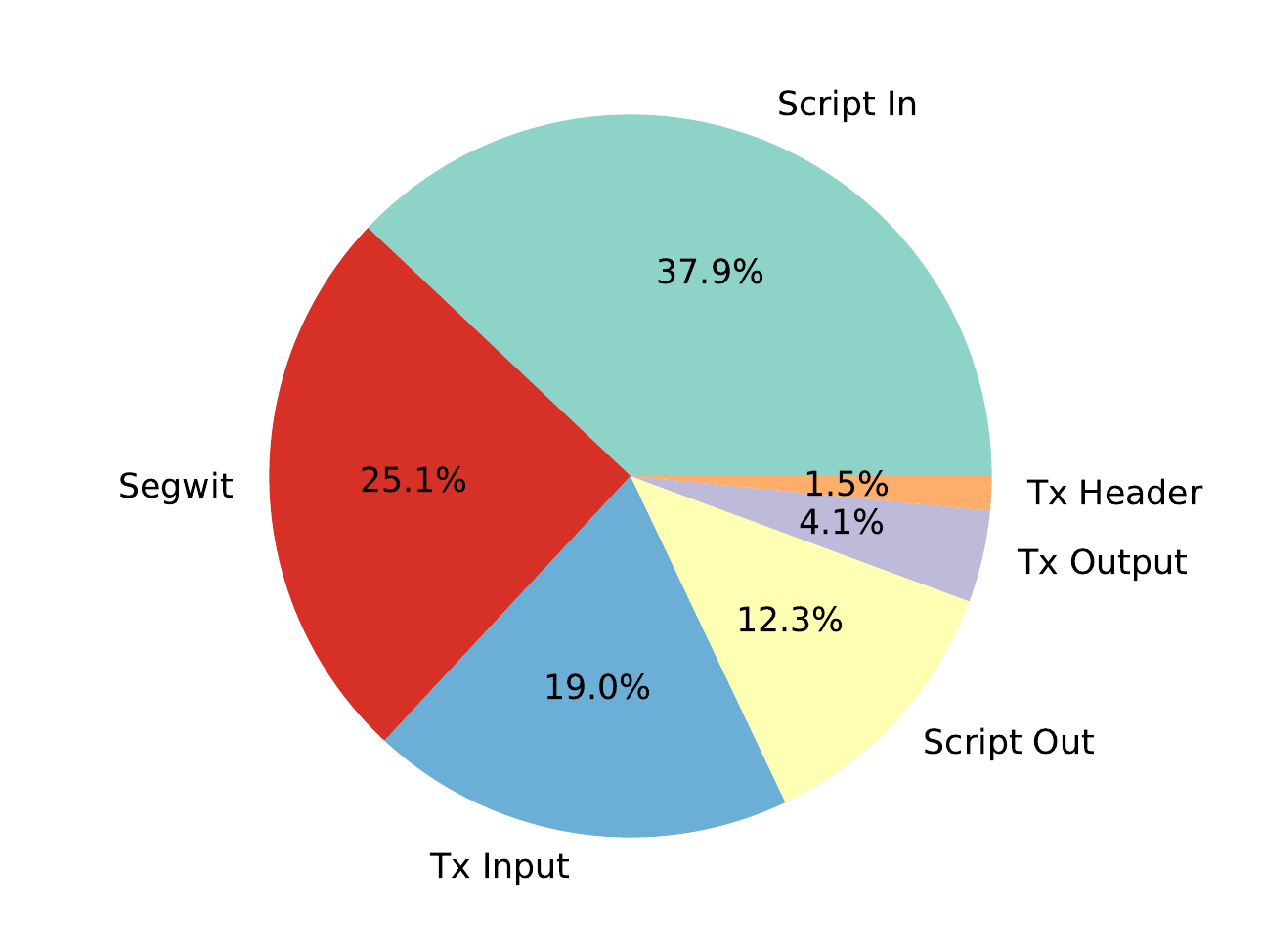}
      \vspace{-3 em}
    \caption{Distribution of data types within transactions from block 481,824 to block 684,816 (i.e., after the appearance of segwit data).}\label{plot_segwit}
\end{figure}

Figures~\ref{plot_no_segwit} show the storage fraction that each data type occupy in a Bitcoin transaction, for transactions between block 0 and block 481,823, i.e., before the introduction of segwit.
Figures~\ref{plot_segwit} provides the same information but for transactions between block 481824 and 684816, that is, after the introduction of segwit.
Our results show that---perhaps unsurprisingly---transaction scripts account for more than 50\% of the size of a transaction.

For example, Figure~\ref{plot_no_segwit} reveals that input scripts and output scripts account for about 62\% and 12\% of a transaction size, respectively.
Figure~\ref{plot_segwit} shows that segregated witnesses reduced the fraction of storage used for input scripts, while adding about 25\% of witness data. The space ratio of other data types remain, as expected, unaffected by segwit.

Given the impact that scripts have on the size of a transaction, we investigate the duplication level of scripts within Bitcoin's ledger. Table~\ref{table:nums}, shows that---in addition to their considerable size---there is large amount of duplicated data among Bitcoin's script. Concretely, Table~\ref{table:nums} shows that about 4,000,000 scripts stored in a TxIn and 40,000,000 scripts stored in a TxOut repeat at least twice. The huge difference between TxIn and TxOut scripts duplication may be caused by unclaimed coins, as well as invalid scripts, or unredeemable transactions (e.g., OP\_RETURN). Nevertheless, our results in Table~\ref{table:nums} shows that in spite of their high duplication, scripts only occupy a small fraction of the Bitcoin ledger storage (approximately 1.5 GB).

\begin{table*}[tbp]
\centering
\scalebox{1.3}{\begin{tabular}{|c|c|c|c|c|} \hline
     \textbf{Script Type} & \textbf{\# of Duplicated Scripts} & \textbf{avg length} & \textbf{Tot. size}& \textbf{Tot. size (Dedup.)}\\ \hline
    TxIn script + segWit & $\sim4\,100\,000$ & 37.6 B & 177 MB & 16 MB\\ \hline
    TxOut script & $\sim40\,500\,000$ & 24.8 B & 1.3 GB & 200 MB\\ \hline
\end{tabular}}
\vspace{0.5 em}
\caption{Duplication of scripts in the Bitcoin blockchain until block 545,000.}
\label{table:nums}
\end{table*}

\subsection{Unused Bytes in Bitcoin Transactions}

Finally, we look at unused bytes in each of a transaction's fields. In particular, we observe that some fields in a Bitcoin transaction have a (large) fixed size; these fields have been designed with the foresight to scale as Bitcoin adoption increases. At present time, the following fields are however largely unused.

\begin{itemize}
    \item \textbf{Transaction headers}:
        \begin{itemize}
            \item \emph{version}: 4 bytes encoding the transaction's data format version. Notice that there only 2 versions available at the time of writing.
            \item \emph{flag}: 2 bytes indicating the presence of witness data.
            \item \emph{lock\_time}: 4 bytes encoding either the block number or the timestamp at which this transaction becomes unlocked. If the transaction does not have a time lock, this field has a default value of \texttt{0xFFFFFFFF}.
        \end{itemize}
    \item \textbf{TxIns}:
        \begin{itemize}
            \item \emph{previous\_output}: 36 bytes are reserved to encode the hash of the transaction referenced by this input (32 bytes) and the index of the specific output to spend in that transaction (4 bytes). Notice that it might suffice to reference the block height and the transaction index at that block height (4 bytes + 2 bytes) along with a variable length integer to encode the specific output to spend in the transaction.
            \item \emph{sequence}: 4 bytes that determine the transaction version as defined by the sender. There only 2 versions available at the time of writing.
        \end{itemize}
    \item \textbf{TxOuts}:
        \begin{itemize}
            \item \emph{value}: 8 bytes to encode the value to be spent (which is a rather large field size).
        \end{itemize}
\end{itemize}

%% file: methodology.tex
\section{Local Storage Optimizations}
\label{sec:meth}

Based on the observations provided in Section~\ref{sec:analysis}, we explore the space of possible solutions that
allow a Bitcoin node to reduce the ledger's footprint on its local storage. Since we are looking at solutions that can
be applied by a node to its local storage without aid by its peers and without modifications to the underlying protocol, we discard techniques that require cooperation among nodes or changes to the underlying protocol (cf. Section~\ref{sec:back}).

The Bitcoin whitepaper originally foresaw two different roles in the Bitcoin blockchain: a \emph{full node} and a \emph{lightweight client}. Lightweight clients cover the needs of basic blockchain users that send or receive transactions but entrust peers to validate transactions. Full nodes were originally designated to contribute to the consensus protocol, by storing and verifying all of the blockchain data, as well as mining new blocks. As Bitcoin grew in adoption and the difficulty of mining increased, a new dedicated \emph{miner} role emerged; these nodes are not meant to store or verify blockchain data, but they are merely optimized to compute PoW solutions.

Beyond these traditional roles, a new type of node is becoming popular: the \emph{archival} node. Such nodes store a full copy of the blockchain for offline auditing and verification purposes but are not interested in real-time verification of transactions.

Given the different types of Bitcoin nodes, one can envision different space-saving strategies that depend on the role of nodes in the ecosystem. For instance, an aggressive data-saving strategy that favors space reduction over the ability to verify all transactions may work for some node types, whereas others might value verifiability of information more than storage savings. In the following, we explore the space of storage-saving options rooted on the observations of Section~\ref{sec:analysis}.

\subsection{Storage Optimization Toolbox}

\vspace{0.5 em}\noindent{\textbf{Pruning based on UTXO lifespan (\prune).}}
The pruning functionality offered by the Bitcoin client is a lossy mechanism and, in case pruned UTXOs are spent, the node must fetch data from the blockchain for verification (thus incurring additional communication overhead).
Our measurements in Section~\ref{sec:analysis} provide solid means to choose a pruning threshold, based on the probability that one of the pruned UTXOs is spent.
According to Figure~\ref{fig:tx_life_cdf}, UTXOs that are older than 42,586 blocks can be pruned, if one tolerates that with 10\% probability one of the pruned UTXO will be spent.

Given the moderate storage costs of hashes, we argue that the block hashes should be kept in this strategy---even for those blocks that are pruned. As discussed in Section~\ref{sec:security}, this ensures that the security of the system is not compromised against sophisticated attacks.

\vspace{0.5 em}\noindent{\textbf{Minimizing Merkle-tree data (\minimize).}} We note that transactions in a block are arranged as leaves of a Merkle tree, so that the root can be used as the authentication token. Rather than keeping all leaves of the tree, one could simply keep the ``co-path'' of transactions with unspent UTXO. Given a
block with $n$ transactions, the co-path to verify an unspent one, amounts to roughly $\log{n}$ nodes of the Merkle tree. Hence, given $k$ unspent transactions in a block, the amount of tree nodes to be stored is $k\cdot\log{n}$. Given that a full binary tree has $n-1$ internal nodes, the strategy of keeping the co-path
of unspent transactions reduces storage as long as $k<n\slash\log{n}$.

Notice that approximately 16\% of the blocks have no UTXOs and can therefore be removed without any penalty.
For those blocks, no intermediate information about the Merkle root/tree must be stored.
Nodes can simply store the block hash as a means to compute the longest chain.

\vspace{0.5 em}\noindent{\textbf{Slack space reduction (\slack).}} As shown in Section~\ref{sec:analysis}, there is considerable room to remove the slack space in several transaction header fields, namely:
\begin{itemize}
             \item \emph{version}: Since there are currently only 2 transaction versions, we can limit the size of this field to 1 bit instead of 4 bytes.
             \item \emph{flag}: Given that this field is only used to indicate the presence of witness data, we can also limit its size to 1 bit. 
             \item \emph{lock\_time}: One can adjust this field as follows: if the transaction does not have a time lock, we include a bit flag. Otherwise, we
             leave the \emph{lock\_field} set time.
         \end{itemize}

Similar techniques can also be applied to the transaction input fields, namely:
         \begin{itemize}
             \item \emph{previous\_output}: One can adjust this field to use a combination of block height and referenced transaction index
             at that block height (4 bytes + 2 bytes) instead of the 32 bytes hash of the transaction, and a variable length integer to encode the specific output
             to spend in the transaction.
             \item \emph{sequence}: Since there are currently only 2 transaction versions, we can limit the size of this field to 1 bit instead of 4 bytes.
         \end{itemize}

Moreover, one can also optimize slack space in the transaction outputs by modifying the current 8-byte \emph{value} field to accommodate for a variable length integer field.%

Finally, one can, in theory, deduplicate existing scripts by implementing a key-value store for efficient script storage. In particular,
given a duplicated script, one can store it in the key-value store (KVS), indexed by its hash, and later replace the script with its hash in every
transaction when the script appears. The local node would then fetch scripts from the key-value store, any time the transaction needs to be verified. However, our experiments show that such a strategy would result in considerable I/O
(to access the KVS) only to result in modest storage savings. Namely, as shown in Table~\ref{table:nums}, scripts (that could be deduplicated) occupy approximately 1.5 GB of storage; replacing such scripts with KVS pointers would only yield a modest
saving of 1.2 GB (0.3\% of the total Bitcoin ledger storage).

\begin{table*}[htbp!]\centering \caption{Evaluation of the various storage-saving strategies.}
\centering
\begin{threeparttable}
\scalebox{1.2}{\begin{tabular}{|c|c|c|c|}
 \hline
 \hline
\multicolumn{2}{|c|}{\textbf{Storage Strategy}} & \textbf{Storage overhead (GB)} & \textbf{Storage reduction (\%)}\\
            \hline
 \multicolumn{2}{|c|}{Full-ledger storage (baseline)}  & 371.4 GB & 0\% \\
 \hline
 \multirow{7}{*}{\rotatebox[origin=c]{90}{\parbox[t]{1.5cm}{\centering \textbf{Standard Compression}}}} & snappy & 335.4 GB & 9.7\%\\ \cline{2-4}
 & lzop & 325.3 GB & 12.41\% \\ \cline{2-4}
 & lz4 & 318.1 GB & 14.35\% \\ \cline{2-4}
 & bzip2 & 302.8 GB & 18.47 \% \\ \cline{2-4}
 & gzip & 300.9 GB & 18.98\% \\ \cline{2-4}
 & zstd & 294.9 GB & 20.59\%\\ \cline{2-4}
 & lzma & 279.6 GB & 24.71\%\\
 \hline
 \multirow{7}{*}{\rotatebox[origin=c]{90}{\parbox[t]{1.5cm}{\centering \textbf{Storage-efficient Modes}}}} & \multicolumn{1}{|c|}{\prune}     &   51.20 GB & 86.22\%\\\cline{2-4}
& \multicolumn{1}{|c|}{\slack{}}    &  265.1 GB  &  28.62\% \\\cline{2-4}
 & \multicolumn{1}{|c|}{\minimize}    &  54.4 GB  &  85.3\%\\\cline{2-4}
 & \multicolumn{1}{|c|}{\minimize{} + \prune}    & 16.5 GB & 95.56\%\\\cline{2-4}

 & \multicolumn{1}{|c|}{\minimize{}  + \slack{}} &  50.6 GB  &  86.37\% \\\cline{2-4}
 & \multicolumn{1}{|c|}{\prune{}  + \slack{}} &  42.4 GB &88.58\% \\\cline{2-4}
 & \multicolumn{1}{|c|}{\prune{} + \minimize{}  + \slack{}}&   15.2 GB   &  95.90\%\\\cline{2-4}
 \hline
 \hline
\end{tabular}}
\end{threeparttable}
\label{table:comparison}
\end{table*}

\subsection{Storage-efficient modes}\label{sec:security}

We now analyze the security and performance tradeoffs of the various storage-saving strategies discussed above. In particular, for each strategy, we analyze the
storage savings and the security implications for a node adopting a particular strategy, in comparison with a full node---one that stores the full blockchain. In terms of
security, we focus on whether a node can validate all transactions or trace back a coin expenditure throughout the entire ledger.

In our analysis, we distinguish between ``lossy'' and ``lossless'' strategies. A lossless strategy, as the name suggests, does not imply any information loss
 compared to the option of storing the full blockchain
but might imply a computational penalty (see discussion below) when verifying transactions. Differently, a lossy strategy incurs in loss of
information when compared to storing the full blockchain
and might incur a communication penalty when verifying transactions---as missing information must be fetched from the network.

Notice that \slack{} is a lossless strategy and, as such, have no impact on security. That is, a node using \slack{} preserve the
same ability of a full node of verifying transactions and tracing back coins.

On the other hand, \minimize{} and \prune{} are lossy strategies.
\prune{} does not allow to verify transactions that include those UTXOs that have been removed. Moreover, both strategies do not allow nodes to trace
coin expenditure, since some information (e.g., transactions confirmed in blocks prior to the pruning threshold in \prune{} or transactions for which the UTXOs are spent in \minimize) has been removed. Hence, a node that uses \minimize{} and/or \prune{} may need to contact a full or archival node  to obtain the missing data when a transactions with a pruned UTXO must be verified or a coin must be traced back.
Nevertheless, we argue that \minimize{} and/or \prune{} do not affect the ability of a node to verify transactions including UTXOs that have not been deleted. In particular, even if not all UTXOs are kept, a node that uses either strategy still keeps the block hashes. Thus, a fraudulent transaction would be considered as valid only if the adversary is able to find a hash collision.
In case of \prune, a proper choice of the pruning threshold is particularly important. One needs to set it high enough so that the vast majority of UTXOs can be verified locally (without the need to fetch data from
other peers) but low enough not to occupy large storage space.

Notice that these strategies can be combined with each other to increase the utility of the nodes---given the role they envision to take in a blockchain (i.e., active verifier, or passive archival node). In particular,
archival nodes can easily adopt \slack, while active verifiers could use a combination \prune{} and \minimize, optionally with \slack.

%% file: evaluation.tex
\section{Evaluation}
\label{sec:eval}
In this section, we evaluate the effectiveness of the strategies discussed in Section~\ref{sec:meth}. To do so, we created a software tool in Python that can be executed locally to estimate the storage footprint of the Bitcoin ledger, given any combination of the storage-saving strategies we devise.
The tool takes as input any combination of the storage-saving strategies of the previous section.
Subsequently, it parses the Bitcoin ledger and outputs the corresponding storage footprint needed to store the ledger.
As mentioned earlier, our tool adapts two existing open-source parsers, namely \texttt{bitcoin-blockchain-parser} \cite{py_parser} and \texttt{bitcoin-tools} \cite{btctools}, both written in Python.

We conducted our experiments on a machine equipped with an Intel\textsuperscript\textregistered Xeon\textsuperscript\textregistered CPU E-2176G @ 3.70GHz and 128GB DDR4 RAM.

\vspace{1 em}\noindent\textbf{Storage savings: }We compare the performance of our space-saving strategies against two different baselines: the full current ledger, and a ledger compressed with standard compression algorithms, namely:

\begin{itemize}
    \item \texttt{bzip2}: lossless compression using the Burrows-Wheeler block sorting text compression algorithm, and Huffman coding.
    \item \texttt{gzip}: lossless compression using Lempel-Ziv coding (LZ77).
    \item \texttt{lzma}: lossless compression using a dictionary compression scheme similar to Lempel-Ziv coding (LZ77).
    \item \texttt{lzop}: lossless compression using the Lempel–Ziv–Oberhumer (LZO) algorithm.
    \item \texttt{lz4}: lossless data compression algorithm focused on compression and decompression speed. Based on Lempel-Ziv coding (LZ77).
    \item \texttt{snappy}: lossless compression algorithm developed by Google. It does not aim for maximum compression, or compatibility with any other compression library; instead, it aims for very high speeds and reasonable compression \cite{snappy}.
    \item \texttt{zstd}: lossless compression algorithm developed by Facebook. It is a fast lossless compression algorithm, targeting real-time compression and better compression ratios \cite{zstd}.
\end{itemize}

To apply standard compression to the Bitcoin ledger, we first created a single non-compressed archive file of the ledger using \texttt{tar}, then fed it to the compression algorithm. 

We note that compressing the ledger typically entails trading-off data saving for performance.
In particular, one could compress the whole ledger to obtain the best savings in terms of space; however, verifying a transaction would require de-compressing the ledger and that may incur considerable delays.
Alternatively, one could compress one block at a time: this is likely to save less space but, at the same time, it is likely to perform better when verifying a transaction since one has to de-compress only some blocks. Our measurements, however, show that even the best compression strategy (i.e., compressing the whole ledger) only yields modest results. For instance, among all studied compression algorithms, \texttt{lzma} achieves the best storage saving and only results in around a 24\% storage saving. When compared to \slack, the latter strategy achieves a 5\% increase (i.e., 29\%) in storage savings (cf. Table~\ref{table:comparison}) while ensuring zero information loss from the Bitcoin blockchain; it also exhibits significant reductions in computational load required for compression/de-compression.

As shown in Table~\ref{table:comparison}, other strategies outlined in Section~\ref{sec:meth} achieve much higher storage efficiency---while requiring less computational overhead. For instance, \minimize{} results in 94\% storage savings and \prune{} can achieve a savings of 86\% while ensuring that 90\% of all UTXOs can be verified with the local data storage (cf. Section~\ref{sec:analysis}).

Various combinations of the aforementioned strategies seem to be also very effective. For instance, the combination of \slack{} and \prune{} results in 88.58\% storage savings, while a straightforward combination of \minimize{} and \prune{} results in up to 95.56\% storage savings.
Finally, the combination of \prune{}, \minimize{}, and \slack{} can result in huge storage savings, up to 95.90\%. That is, the reliance on \slack{} would only result in mediocre storage gains when \prune{} and \minimize{} are being used.

\vspace{1 em}\noindent\textbf{Computational  overhead: }In order to assess the computational overhead of the proposed strategies, we additionally measured the time it took to our tool to run \minimize{} and \slack{} over each block between block 682,807 and block 682,816 of the Bitcoin blockchain. On average. \minimize{} took 2.1ms (std-dev 0.9ms) whereas \slack{} took 102.9ms (std-dev 58.7ms). We conclude that both strategies incur in negligible processing overhead. We did not evaluate the overhead incurred in \prune{} since it merely requires the peer to \emph{delete} a given block.

%% file: conclusions.tex
\section{Concluding Remarks}
\label{sec:concl}
In this paper, we tackled an often overlooked issue of today's blockchains---reducing the ledger size.

We observe that current blockchains do not employ storage-efficient strategies, their ledger features considerable redundancy, and they rarely utilize all the data stored in the ledger for daily operations.
We apply our reasoning to Bitcoin's data storage, and show, by means of empirical measurements, that the ledger storage can be considerably reduced without modifying the underlying consensus protocol nor affecting the security of the verification process. To this end, we adapted the blockchain parser from~\cite{py_parser} and measured the storage footprint of various local strategies that can be directly employed at client-side by full nodes in Bitcoin.

Our evaluation results show that standard compression algorithms are not effective in capturing the intrinsic nature of Bitcoin's ledger and only result in up to 24\% storage savings. On the other hand, more fine-grained lossless compression strategies---those that specifically target unused bytes or duplication in the blockchain---prove to be more effective and could lead to storage savings up to 29\%.

In terms of lossy strategies, our results also show that ledger pruning---a popular strategy to reduce the local storage footprint of nodes---would require at least 51~GB of storage space in Bitcoin in order to process the vast majority of those UTXOs in circulation. Besides laying the grounds that govern the effective choice of a suitable pruning threshold, we also show that pruning can be combined with other lightweight strategies to incur a lower storage footprint, as low as 15.2 GB without incurring significant computational overhead on nodes.

We plan to release our parser as open-source to better aid the community in estimating the actual storage needs of Bitcoin nodes as the ledger grows in size. Finally, we stress that our observations and results are not restricted to the Bitcoin blockchain and equally apply to the myriad of altcoins (or forks of the Bitcoin blockchain) that are currently deployed (e.g., Dogecoin, Bitcoin Cash, Litecoin, Monacoin).